\documentclass[final,5p,twocolumn]{elsarticle}
\usepackage{amssymb}
\usepackage{amsmath}
\usepackage{amsfonts}
\usepackage{bm}
\usepackage{enumerate}
\usepackage{color}
\usepackage{graphicx}

\begin{document}

\begin{frontmatter}



%
\title{Unitarity of quantum tunneling decay for an analytical exact \\non-Hermitian resonant-state approach}

\author[1]{Gast\'on Garc\'{\i}a-Calder\'on\corref{cor1}}
\ead{gaston@fisica.unam.mx}
\author[2]{Roberto Romo}
\cortext[cor1]{Corresponding author}
\address[1]{Instituto de F\'{\i}sica, Universidad Nacional Aut\'onoma de M\'exico, 04510 Ciudad de M\'exico, M\'exico}
\address[2]{Facultad de Ciencias, Universidad Aut\'onoma de Baja California, 22800 Ensenada, Baja California, Mexico}

\begin{abstract}
By using an analytical exact non-Hermitian formalism for quantum tunneling decay that involves the expansion of the decaying wave function as a linear combination  of resonant states and transient functions associated with the complex poles of the outgoing Green's function to the problem, it is shown that the integrated decaying probability density in the whole space satisfies unitarity at each value of time.
\end{abstract}

\end{frontmatter}
\section{Introduction}

Quantum tunneling decay refers to an open system that involves the energy continuum where a particle initially confined by a potential barrier decays to the outside by tunneling. Here we want to address the issue of unitarity of non-Hermitian Hamiltonians involving real potentials of arbitrary shape that vanish exactly beyond a distance.

In conventional quantum mechanics the time evolution operator is unitary because the Hamiltonian is Hermitian. As a consequence, the eigenvalues are real and the norm of the evolving state remains constant in time, a fact that reflects the physical requirement of conservation of probability. For quantum tunneling, the decaying wave solution may be written as an expansion in terms of the continuum wave functions to the problem and unitarity follows from Dirac normalization of these functions. However, the exact calculation of the decaying wave function requires of a full numerical calculation which, in addition to be cumbersome and computationally time consuming, it provides no physical insight on the dynamics of the decay process.

The above consideration on unitarity seems to be contradiction with approaches involving non-Hermitian Hamiltonians, originated in the early days of quantum mechanics in the work by Gamow to describe  $\alpha$-decay in radioactive nuclei \cite{gamow28a,gamow28b}. He considered an analytical non-Hermitian approach that on physical grounds follows by imposing outgoing boundary conditions to the solutions of the time-dependent Shcr\"odinger equation to the problem. This leads to complex energy eigenvalues and eigenfunctions with a divergent norm and to a decaying probability density that decreases exponentially with time which seems to violate unitarity, but that on the other hand, led to the analytical expression of the \textit{exponential decay law} \cite{gamow28a,gamow28b}, which characterizes a decaying regime that has been amply verified experimentally over the years in a variety of tunneling decaying systems \cite{gamow49,sakaki87,jochim11}. However, since non-Hermitian approaches lie outside the conventional framework of quantum mechanics there is widespread view that considers them as a simplified, phenomenological and a nonfundamental description of the tunneling decay process \cite{bender07,barton63}.

The 1939 work by Siegert \cite{siegert39} contribute to relate the notion of resonance in scattering and decay processes, which led to identify the complex energy eigenvalues with the complex poles of the scattering matrix and after the work by Peierls \cite{peierls55}, with the complex poles of the propagator. Since the imaginary part of the complex energy determines both the width of the resonance and the decay rate in the time evolution of decay, these states are commonly named resonant states \cite{peierls59,rosenfeld61,gcp76} even if the resonances overlap. One finds in the literature other names for these states which refer to related approaches, as Gamow states \cite{madrid12}, quasinormal states \cite{leung98}, Siegert states \cite{tolstikhin08}, Siegert pseudostates \cite{tolstikhin98} and Gamow-Siegert states \cite{rosas08b}.

It may be of interest to point out that expansions involving the complex poles of the scattering matrix $S(k)$ as occur, for example, for the $s$-wave continuum wave solution $\phi(k,r)=(i/2)[\exp(-ikr) -S(k)\exp(ikr)]$ \cite{peierls59,rosenfeld61}, are Hermitian because  these functions satisfy the corresponding Schr\"odinger equation to the problem with real complex eigenvalues, i.e., $[k^2-H]\Phi^+(k,r)=0$ and form a complete set of functions involving Dirac normalization. As pointed out lucidly by Moiseyev, what determines the non-Hermitian character of the Hamiltonian is that the functions upon which the Hamiltonian acts are not in the  Hermitian sector of the domain of the Hamiltonian \cite{moiseyev11}. This is indeed the case for outgoing boundary conditions on the solutions to the Shr\"odinger equation which yield $[k^2_n-H]u_n(r)=0$, where $k^2_n=\mathcal{E}_n-i\Gamma_n/2$ and are not normalized in the usual sense.

It is worth mentioning that the second half of the last century witnessed the prediction of deviations from exponential decay both at ultrashort and very long times compared with the lifetime to the system \cite{khalfin58,khalfin68,chiu77}, which have been verified experimentally \cite{raizen97,raizen01,monk06}, and hence indicate that the tunneling decay process is a much more complex phenomenon
than previously envisaged. The above theoretical work is based on the notion of survival probability, which yields the probability that the evolving decaying state remains in the initial state. Since the initial state is confined within the internal region of the interaction potential, the divergent character of the decaying solution with distance does not arise in these treatments. Clearly, however, the survival probability  is unsuitable to provide a description of the propagation of the decaying wave solution outside the interaction region of the potential and hence it misses interesting theoretical findings, as for example, that the decaying probability density beyond a distance decays purely in an non exponential fashion \cite{torrontegui09,gcmv13}.

An important feature of the decaying wave function in the work by Gamow, Khalfin and  most of subsequent work \cite{gcp76,mgcm05}, is that only involves \textit{proper} complex poles, which on complex $k$ plane are those complex poles located on the corresponding fourth quadrant, such that the real parts are larger than the imaginary ones. As discussed briefly below, all these approaches lead to a non-unitary description.

In this work we consider the expansion of the decaying wave function by Garc\'{\i}a-Calder\'on and coworkers \cite{gc10} that involves the full set of resonant states (both proper and nonproper) to the system and transient functions which depend on the complex poles of the outgoing Green's function to the problem, to show analytically that each term of the  expansion is quadratically integrable for each value of time and hence that unitarity is fulfilled in this approach.

It is worth pointing out that the transient function, which is proportional to the complex error function \cite{abramowitzchap7,faddeyeva61}, appears to be a relevant quantity for describing transient phenomena in quantum mechanics \cite{cgcm09}. This function was considered by Moshinsky in a schematic model for scattering and desintegration involving complex poles of the scattering matrix \cite{moshinsky51} and to predict the  phenomenon of \textit{diffraction in time}  \cite{moshinsky52} which has been verified experimentally \cite{dalibard96}. The study of \textit{diffraction in time}, that refers to free time evolution,  was extended by  Garc\'{\i}a-Calder\'on and Rubio for finite range  potentials of arbitrary shape  by considering an expansion of the time-dependent solution in terms of resonant states and transient functions \cite{gcr97}.  The properties of these functions have also been considered by Faddeyeva and Terent'ev \cite{faddeyeva61}. A consequence of the above is that some authors refer to the these functions as Moshinsky functions and others as Faddeyeva functions. We presently believe that it might be more appropriate to refer to them simply as transient functions, which emphasize that they refer to transient non stationary processes. For potentials that vanish beyond a distance transient functions depend on the poles of the scattering matrix or equivalently of the corresponding outgoing Green's function of the problem. We should mention that some authors have studied transients in scattering and decay without referring to the notion of resonant state  \cite{rosenfeld65,nussenzveig92} or by considering functions that actually are proportional to the resonant states of the problem \cite{dijk99}.

The  full resonant state  expansion of the decaying wave solution involving transient functions has been used extensively by Garc\'{\i}a-Calder\'on and coworkers to investigate in a unified framework the short, exponential and long time behaviors of the survival and nonescape probabilities \cite{gc10,cgc12,gcmv07}, and to explore the conditions to observe the deviation from exponential decay at  long times in artificial quantum systems \cite{gcv06,gcr16}, and more recently regarding fundamental issues of quantum mechanics \cite{gcch17, gcv19}. The present work may be inscribed in this line of research.

For the sake of completeness, we would like to mention the growing interest in recent years on  non-Hermitian approaches that in general are not concerned with tunneling decay, as the so called PT symmetry \cite{muga05,mostafazadeh09c,bender15} and on exceptional points, which correspond to complex double poles of the outgoing Green's function \cite{barrios18}.

The paper is organized as follows. In Section II we discuss the lack of unitarity in approaches that involve only \textit{proper} complex energy  poles. Section III reviews some relevant properties of resonant states, which involve the complex poles of the propagator and the resonant states to the problem and discusses the time evolution of the decaying solution. In Sec. IV we analyze the asymptotic behavior of the decaying solution as a function of distance for fixed values of time. In Sec. V, we exemplify our findings using an exactly solvable model, and finally, Section VI presents some concluding remarks.

\section{Lack of unitarity of resonant expansions of the decaying wave solution for \textit{proper} complex poles }

We consider for this work a central potential $V(r)$ having a barrier from which a particle initially confined within the interaction region of the potential escapes to the outside by tunneling. We assume, based on physical grounds, that the potential vanishes exactly beyond a distance, namely, $V(r)=0$ for $r>a$. For the sake of simplicity and without loss of generality we restrict the discussion to zero angular momentum. Also, we set natural units $\hbar=2m=1$.

The approach by Gamow to describe $\alpha$-decay was to replace the many-body nuclear potential by a single particle potential and proceed to obtain the decaying wave function as the solution to the time-dependent Schr\"odinger equation of the problem obeying, on physical grounds, outgoing boundary conditions. As is well known, that lead to a set of discrete complex wavenumbers $\kappa_n=\alpha_n-i\beta_n$, with $\alpha_n> \beta_n$, and hence to complex energy eigenvalues $\kappa^2_n=E_n=\mathcal{E}_n-i\Gamma_n/2$. The outgoing boundary condition was the original theoretical contribution that left Gamow's approach outside the conventional framework of quantum mechanics. Indeed, Gamow's propagating solution may be written as \cite{gamow49},
\begin{equation}
\Psi_{G}(r,t)= [e^{i\alpha_n r}e^{-i\mathcal{E}_n t}]e^{\beta r}e^{-\Gamma_n t/2},  \quad r \geq a,
\label{b6x}
\end{equation}
which shows analytically, as is well known, that for a given time $\Psi_{G}(r,t)$ diverges with distance and hence it does not satisfy unitarity, namely,
\begin{equation}
\int_a^{\infty}|\Psi_{G}(r,t)|^2 \,dr \to \infty.
\label{ge}
\end{equation}

A convenient form to discuss resonant expansions for the decaying wave function was discussed in Ref. \cite{gcp76}. Here, for completeness of the presentation we highlight the main steps. The decaying wave function $\Psi(r,t)$ may be written as an integral involving the retarded Green's function $g(r,r';t)$ as,
\begin{equation}
\Psi(r,t)=\int_0^a {\! g(r,r^\prime;t)\Psi(r',0)\,\mathrm{d}r^\prime},
\label{1s}
\end{equation}
where $\Psi(r,0)$ stands for an arbitrary initial state which is confined within the internal interaction region.
The retarded time-dependent Green's function $g(r,r';t)$ is the relevant quantity to study the time evolution of the initial state for $t > 0$ and  may be evaluated  by a Laplace transformation into the complex wave number plane $k$,
\begin{equation}
g(r,r';t)={i \over 2 \pi} \int_{C_0} G^+(r,r\,';k) e^{-i k^2t} \,2kdk,
\label{s2}
\end{equation}
where $C_0$ represents a Bromwich contour along the first quadrant on the $k$ plane. It is well known that for potentials that vanish exactly after a distance, the outgoing Green' function to the problem $ G^+(r,r\,';k)$ has an infinite number of complex poles distributed on the third and fourth quadrants symmetrically with respect to the ${\rm Im}\,k$-axis \cite{newtonchap12}. Since our description refers to the energy continuum, we assume for the sake of simplicity and without loss of generality, that there are no bound nor antibound poles. One may then  close the integration contour in the $k$ plane to pick up the contribution of the complex poles $\kappa_n$ using the theorem of residues.  The factor $\exp(-ik^2t)$ converges only in the second and fourth quadrants of the $k$ plane and hence the above procedure leads to a description involving only \textit{proper} complex poles.
The resonant state functions $u_n$ follow from the residues $\rho_n(r,r')$ at the complex poles of the outgoing Green's function to the problem $G^+(r,r\,';k)$ \cite{gcp76,gcr97,gc10}, namely, $\rho_n=u_n(r)u_n(r')/2\kappa_n$, which sets the normalization condition,
\begin{equation}
\int_0^a u_n^2(r) dr + i\frac{u_n^2(a)}{2\kappa_n}=1.
\label{r4}
\end{equation}

Aypical expression for the decaying wave function reads \cite{gcp76,mgcm05},
\begin{equation}
\Psi(r,t)= \sum_{n=1}^\infty C_nu_n(r)e^{-i\mathcal{E}_n t}e^{-\Gamma_n t/2} +Z(r,t),
\label{gcp}
\end{equation}
where the coefficients $C_n$ refer to the overlap of the initial state $\Psi(r,0)$ with the resonance state $u_n(r)$, and $Z(r,t)$ stands for an integral term that accounts for the nonexponential contributions to decay.
An essentially similar expansion of the decaying wave function  $\Psi(r,t)$  (\ref{gcp}) may also be obtained using  the rigged-Hilbert space approach \cite{mgcm05,madrid12}.

The relevant point here is that for $r>a$, each resonance term in (\ref{gcp}), behaves as in (\ref{b6x}) and hence it  diverges with distance. Consequently, as in  Gamow's case, the integrated probability density is manifestly non-unitary. In fact, notice that Gamow's solution follows from (\ref{gcp}) by considering just a single resonant term with unity coefficient and omitting the nonexponential term.

The exponential grow with distance of the propagating resonant solutions together with its exponential time decrease  has led to a number of authors to argue that the physical understanding of unitarity must consist of an interrelated consideration of both the space and time features of the decaying solution \cite{gamow49,rosenfeld61,hatano09,madrid12}.

\section{Expansion of the decaying wave solution in terms of the full set of resonant states and transient functions}

Let us now briefly discuss the expansion of the decaying wave solution involving the full set of complex poles, which consist of both \textit{proper} and \textit{non proper} poles, where the \textit{non proper} poles correspond mainly to those located on the third quadrant of the complex $k$ plane.

One may instead of closing  the Bromwich contour along a path of the complex $k$ plane to obtain the resonant expansion (\ref{gcp}), to close the contour $C_0$ in (\ref{s2}) in a different form to write the retarded time-dependent Green's function $g(r,r';t)$ as \cite{gc10,gcmv12},
\begin{equation}
g(r,r';t)={i \over 2 \pi} \int_{-\infty}^{\infty} G^+(r,r\,';k) {\rm e}^{-i k^2t} \,2kdk.
\label{74}
\end{equation}
The evaluation of $g(r,r';t)$ may be obtained by noticing that $G^+(r,r\,';k)$ itself may be expanded in terms of the full set of resonance states \cite{gc10,gcmv12}. This indeed corresponds to a very different procedure than those discussed in the previous section. It requires to study the behavior of $G^+(r,r\,';k)$ as $k \rightarrow \infty$ along all directions of the complex $k$ plane and it may be proved that it leads to a convergent resonant expansion for $G^+(r,r\,';k)$ provided $r$ and $r'$ are smaller than the boundary radius $a$ or one of these is at  $a$ whereas the other remains smaller than $a$ \cite{gcb79}. Denoting this by the notation $(r,r')^{\dagger} \leq a$, one may write  the outgoing Green's function to the problem as the expansion \cite{gc10,gcmv12},
\begin{equation}
G^+(r,r\,';k) = \frac{1}{2k}\sum_{n=-\infty}^{\infty} \frac {u_n(r)u_n(r\,')}{k-\kappa_n}
\quad  (r,r')^{\dagger} \leq a,
\label{9xa}
\end{equation}
where we emphasize that the sum runs over the full set of complex poles  $\kappa_n$ and $\kappa_{-n}$, located respectively on the third and fourth quadrants of the $k$ plane. The resonance states $u_{-n}(r)$ and complex poles $\kappa_{-n}$ located on the third quadrant of the $k$ plane are related to those located on the fourth quadrant by symmetry relations that follow from time reversal invariance: $\kappa_{-n}= -\kappa_n^*$ and $u_{-n}(r) = u^{*}_n(r)$ \cite{rosenfeld61,gc10}.
Substitution of (\ref{9xa}) into (\ref{74})  and then into (\ref{1s}) allow us to write the time-dependent decaying wave function $\Psi(r,t)$ for $r\leq a$.

For $r > a$ we may express of $G^+(r,r\,';k)$ in terms of the regular and irregular solutions to the Schr\"odinger equation \cite{newtonchap12}, to write the identity,
\begin{equation}
G^+(r,r\,';k)=G^+(r',a;k) e^{ik(r-a)},\quad r'<a, \quad r\geq a.
\label{9yzz}
\end{equation}
Then using (\ref{9xa}) one may expand  $G^+(r',a;k)$ in terms of the full set of resonance states and substitute the resulting expression into (\ref{74}) and then into (\ref{1s}), to obtain the resonance expansion of the decaying wave function for $r'<a$ and $r \geq a$. Hence, we may finally write the decaying wave functions as \cite{gc10,gcmv12},
\begin{equation}
\Psi(r,t)=
\left \{ \begin{array}{cc}
\Psi_{in}(r,t), & \quad  r \leq a, \\[.4cm]
\Psi_{ex}(r,t), & \quad r \geq a,
\end{array}
\label{b6}
\right.
\end{equation}
where $\Psi_{in}(r,t)$ and $\Psi_{ex}(r,t)$ are given by,
\begin{equation}
\Psi_{in}(r,t)=\sum_{n=-\infty}^{\infty}C_nu_n(r)M(y^\circ_n),  \quad  r \leq a
\label{b6i}
\end{equation}
and
\begin{equation}
\Psi_{ex}(r,t)=\sum_{n=-\infty}^{\infty}C_nu_n(a)M(y_n),  \quad r \geq a,
\label{b6e}
\end{equation}
with the coefficients $C_n$ in the above two expressions given by,
\begin{equation}
C_n=\int_0^a \Psi(r,0) u_n(r) dr.
\label{3c}
\end{equation}
The functions $M(y_n)$ in (\ref{b6e}) are defined as \cite{gcmv12,gc10},
\begin{eqnarray}
M(y_n)&=&\frac{i}{2\pi}\int_{-\infty}^{\infty}\frac{e^{ik(r-a)}e^{-ik^2t}}{k-\kappa_n}dk
\nonumber \\ [.4cm]
&&=\frac{1}{2}{e^{i(r-a)^2/4 t}} \,w(iy_n),
\nonumber \\ [.4cm]
&&=\frac{1}{2}e^{i(r-a)^2/4 t}e^{y^2_n}{\rm erfc}(y_n)
\label{16c}
\end{eqnarray}
with
\begin{equation}
y_n= e^{-i\pi /4}(1/4t)^{1/2}[(r-a)-2 \kappa_nt].
\label{16cd}
\end{equation}
The function $w(z)={\exp}(-z^2)\rm{erfc}(-iz)$ in (\ref{16c}) stands for the complex error function, also named Faddeyeva or  Faddeyeva-Terent'ev function \cite{abramowitzchap7,faddeyeva61}, for which there exist computational tools to calculate it, as Mathematica or following Poppe and Wijers \cite{poppe90}. The argument $y_n^{\circ}$ of the functions $M(y_n^\circ)$ in (\ref{b6i}) is that of $y_n$ with $r=a$, namely,
\begin{equation}
y^\circ_n=-{\rm e}^{-i\pi /4}\kappa_nt^{1/2}.
\label{16ci}
\end{equation}
\subsection{Analysis of the time evolution of the decaying solution}

This subsection briefly discusses both the internal and external decaying wave solutions solutions of the potential.

\subsubsection{Internal resonance solution}

The properties of the decaying solution along the internal interactionn region has been discussed previously
\cite{gc10,gcmv12}. Here for completeness of the discussion we jut recall some relevant expressions for our discussion.
Using the symmetry relations mentioned above, namely, $\kappa_{-n}=\kappa^*_n$ and $u_{-n}(r)=u^*_n(r)$ allows us to write $\Psi_{in}(r,t)$ given by (\ref{b6i}) as,
\begin{equation}
\Psi_{in}(r,t)= \sum_{n=1}^\infty [C_nu_n(r)M(y_n^\circ)+{\bar C}_n^* u_n^*(r)M(y_{-n}^\circ)],
\label{i2}
\end{equation}
where ${\bar C}_n$ is given by $C_n$, defined by (\ref{3c}), with $\Psi(r,0)$ replaced by $\Psi^*(r,0)$.
One then may utilize a property of the transient functions to write $M(y_n^\circ)$ as,
\begin{equation}
M(y_n^\circ) =  e^{-i\kappa_n^2t} -M(-y_n^\circ),
\label{r5}
\end{equation}
which follows provided that $\pi/2 < \arg \,(y_n^\circ) < 3\pi/2$ \cite{moshinsky52,gc10}. This is in fact the case for resonant poles with $\alpha_n > \beta_n$, the so called \textit{proper} resonant poles. The arguments of both $M(-y_n^\circ)$ and $M(y_{-n}^\circ)$, satisfy $-\pi/2 < \arg \,(y_n^\circ) < \pi/2$, and as a consequence they do not exhibit an exponential behavior \cite{moshinsky52,gc10}. As a result, one may write $\Psi_{in}(r,t)$ for $r\leq a$ as,
\begin{equation}
\Psi_{in}(r,t)= \sum_{n=1}^\infty C_nu_n(r)e^{-i\mathcal{E}_n t}e^{-\Gamma_n t/2} + R(r,t),
\label{i2a}
\end{equation}
where $I(r,t)$ accounts for the nonexponential contribution \cite{gc10},
\begin{equation}
R(r,t)= -\sum_{n=1}^\infty [ C_nu_n(r) M(-y_n^\circ)-{\bar C}_n^*u_n^*(r)M(y_{-n}^\circ)].
\label{i4}
\end{equation}
It is well known that the nonexponential term (\ref{i4}) is relevant at ultra short or very long times compared with the lifetime to the system \cite{cgc12,gcmv12}.

The decaying solution $\Psi_{in}(r,t)$,  given by (\ref{i2a}), is the relevant ingredient to calculate the survival
probability $S(t)=|A(t)|^2$ defined previously and the nonescape probability $P(t)$, which yields the probability that at time $t$ the decaying particle remains within the interaction region. These two quantities have been amply discussed using the formalism of resonant states by Garc\'{\i}a-Calder\'on and coworkers in a number of works \cite{gcmv07,gcrv07,gc10,cgcrv11,gcmv12}.

\subsubsection{External resonance solution}

For $r \geq a$, the solution $\Psi_{ex}(r,t)$, given by  (\ref{b6e}),  describes the propagation of the decaying particle along the external region. This has been calculated for a fixed distance as a function of time in Refs. \cite{gcmv12,gcmv13,gc10}. Instead, we provide below an analysis of its properties for a fixed time $t$ as a function of distance.

To discuss the propagation along the external region it is convenient to write (\ref{b6e}), in a similar fashion as for the internal solution, to show explicitly the contributions corresponding to the third quadrant of the $k$ plane,
\begin{equation}
\Psi_{ex}(r,t) =\sum_{n=1}^\infty [ C_nu_n(a) M(y_n)+{\bar C}^*_nu_n^*(a)M(y_{-n})].
\label{i2be}
\end{equation}
The exponential and nonexponential behavior of $\Psi_{ex}(r,t)$ may be obtained in a similar fashion as in the case of $\Psi_{in}(r,t)$ given above, except that in the present case it depends on the overall sign of the argument $y_n$ given by (\ref{16cd}) due to the presence of the term $(r-a)$. To clarify this, is convenient to write explicitly the real and imaginary parts of $y_n$ for a fixed time $t=t_0$, namely,
\begin{eqnarray}
y_n&=&\frac{1}{2(2t_0)^{1/2}}\left[(r-a)-2(\alpha_n-\beta_n)t_0\right]  \nonumber \\ [.4cm]
&&-i\left [(r-a)-2(\alpha_n+\beta_n)t_0\right].
\label{n1}
\end{eqnarray}
One sees that the sign of real part of (\ref{n1}) depends on whether $(r-a)$ is larger or smaller than $2(\alpha_n-\beta_n)t_0$. The case $(r-a) < 2(\alpha_n-\beta_n)t_0$ satisfies $\pi/2 < \arg \,(y_n) < 3\pi/2$ and hence, in a similar fashion as for the internal case, it exhibits an explicit exponential behavior, namely,
\begin{eqnarray}
\Psi_{ex}(r,t)&=& \sum_{n=1}^\infty C_nu_n(a)e^{i\kappa_n(r-a)}e^{-i\mathcal{E}_n t}e^{-\Gamma_n t/2}
\nonumber \\ [.4cm]
&&+J(r,t),
\label{i2ae}
\end{eqnarray}
where the nonexponential term $J(r,t)$  reads,
\begin{equation}
J(r,t)= -\sum_{n=1}^\infty [ C_nu_n(a) M(-y_n)-{\bar C}^*_nu_n^*(a)M(y_{-n})].
\label{i4e}
\end{equation}
\begin{figure}
\begin{center}
\includegraphics[width=1.0\linewidth]{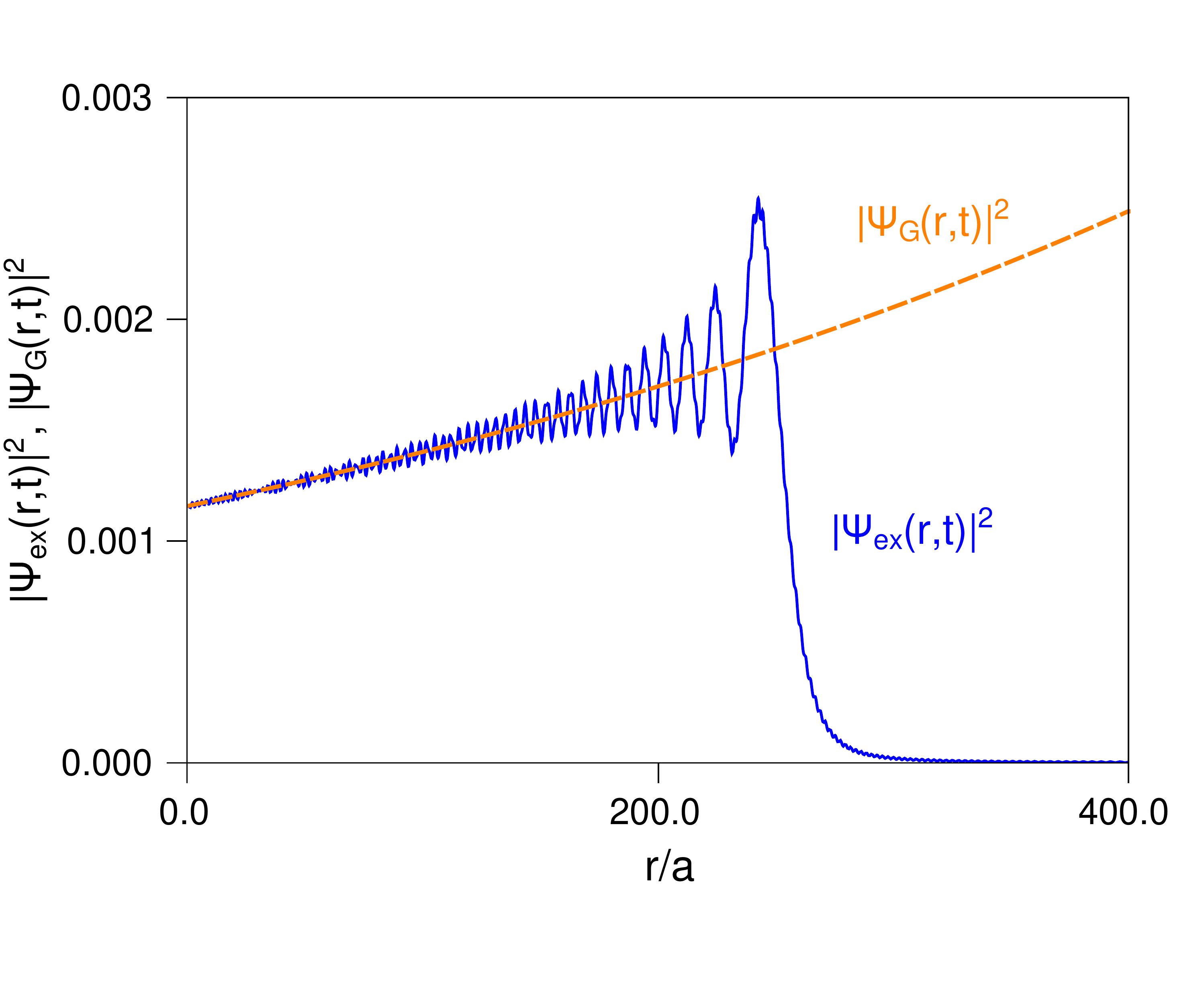}
\caption{Plot of the Gamow solution  $|\Psi_G(r,t)|^2$, given by (\ref{b6x}), and the resonance expansion $|\Psi_{ex}(r,t)|^2$, given by (\ref{b6e}),  as a function of distance along the external region at a fixed time in lifetime units to exhibit the exponential catastrophe of the Gamow solution  and the well behaved wavefront of the resonance expansion. See text.}
\label{figure1}
\end{center}
\end{figure}

On the other hand,  in the case $(r-a) > 2(\alpha_n-\beta_n)t_0$, a decomposition as that given in (\ref{i2ae}) no longer applies, since  the argument $y_n$ satisfies $-\pi/2 < \arg \,(y_n) < \pi/2$, and hence the solution (\ref{i2be}) behaves entirely in a nonexponential fashion.

A consequence of the above considerations is that the propagating solution $\Psi_{ex}(r,t)$ grows exponentially in an oscillatory form until it reaches the value $(r-a)=2 (\alpha_n-\beta_n) t_0$,  and  subsequently it behaves in a nonexponential fashion. In the next Section we show analytically, for a given fixed time $t_0$, that for large values of $r$ the solution given by (\ref{i2be})  goes as $1/r$.

It is worth pointing out, in addition to the above discussion, that by assuming that the initial state $\Psi(r,0)$ is normalized to unity, it follows from the modified closure relation obeyed by resonant states that \cite{gc10,gcmv12},
\begin{equation}
{\rm Re}\sum_{n=1}^\infty \left\{ C_n \bar{C}_n\right\}= 1.
\label{9z}
\end{equation}
Equation (\ref{9z}) indicates that ${\rm Re}\,\{C_n{\bar C}_n\}$ cannot be interpreted as a probability, since in general it is not a positive definite quantity. However, it may be seen as the `strength'  or `weight' of the initial state in the corresponding resonant state.
\begin{figure*} [!tbp]
\centering
\includegraphics[width=14.0 cm,height=9cm]{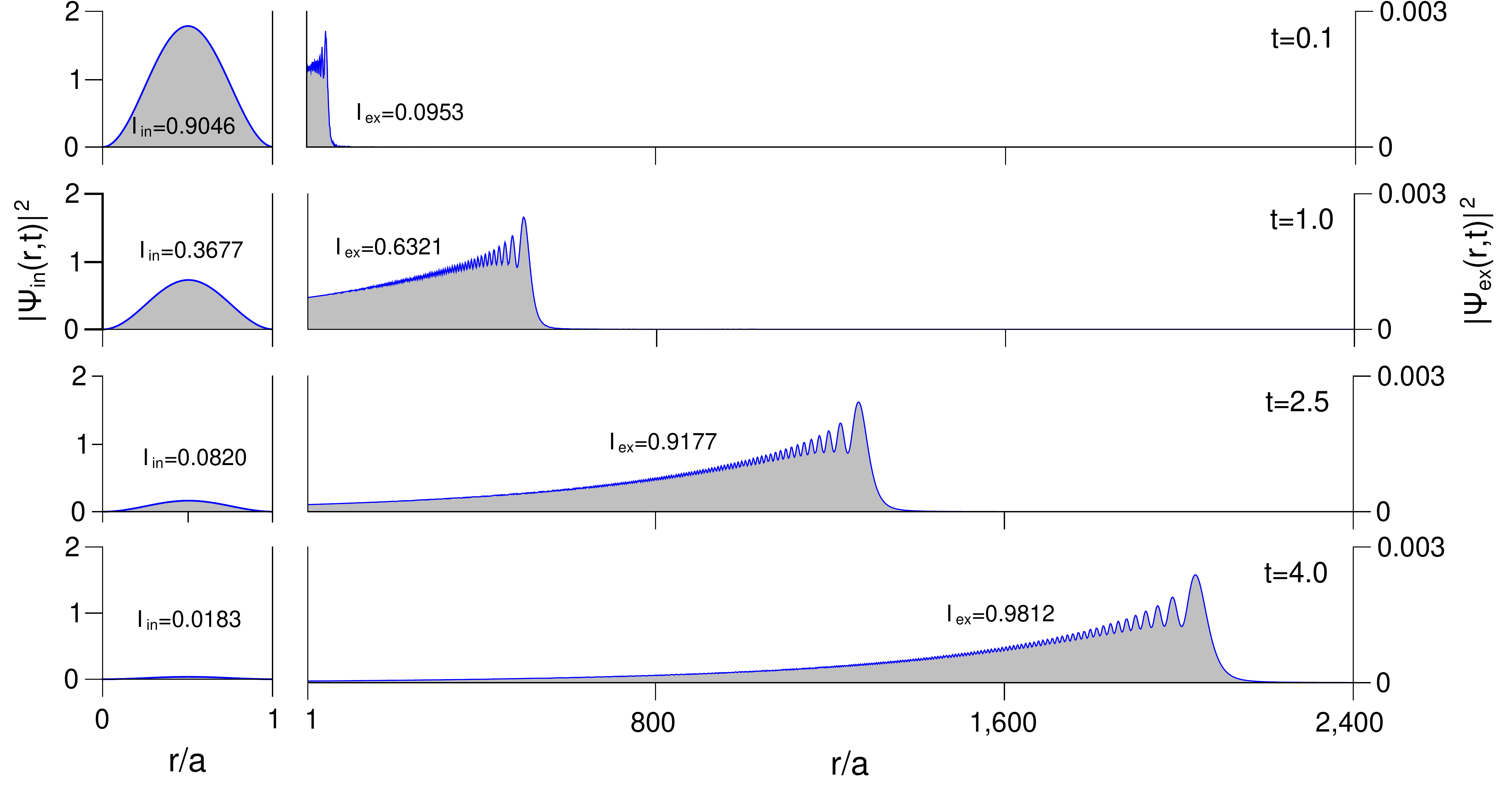}
\caption{Decay and propagation at different fixed times for a normalized quantum box initial state ($q=1$). At each time (in lifetime units), the graph on the left corresponds to the decaying probability density in the internal region of the confining potential, whereas the graph on the right shows the propagating probability density along the external region of the potential. The numerical values of the integrated probability density inside and outside the potential, $I_{in}(t)$ and $I_{ex}(t)$, are indicated in each graph. As may be easily verified, for each value of $t$ the integrated probability density $I_{in}(t)+I_{ex}(t)$, given by (\ref{ut}), yields the value $0.999$, which is  already very close to unity and shows the fulfillment of unitarity. See text.}
\label{figure2}
\end{figure*}
\begin{figure}[htbp!]
\includegraphics[width=3.6in,height=2.5in]{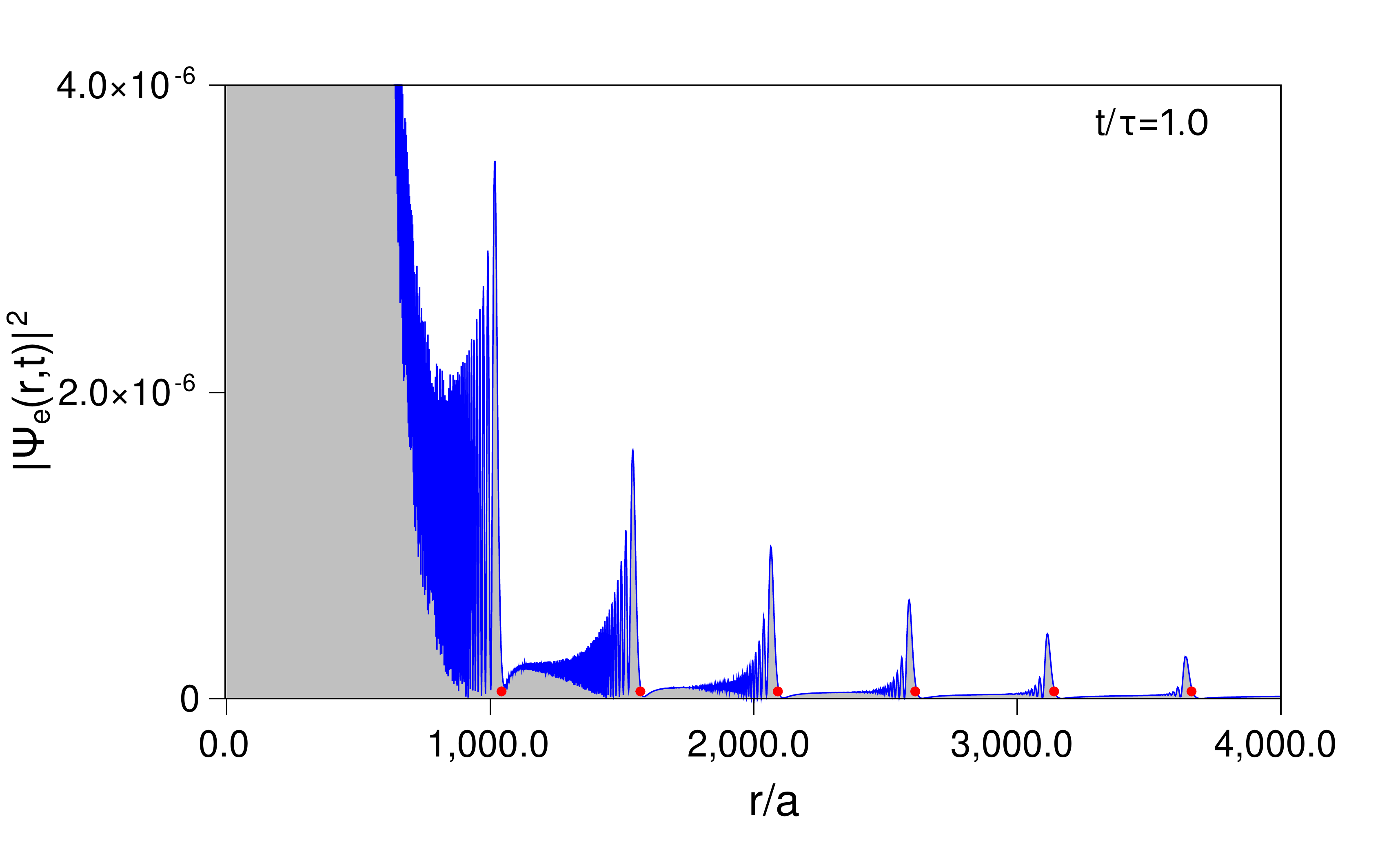}
\caption{ Plot of a zoom of the second graph of Fig. \ref{figure1} extended up to $r/a=4000$, to show the propagating forerunners that arise from the high energy resonance levels of the system. The dots (red) indicate the positions associated with the corresponding resonance velocities. See text.}
\label{figure3}
\end{figure}
\begin{figure*} [!tbp]
\centering
\includegraphics[width=14.0 cm,height=9cm]{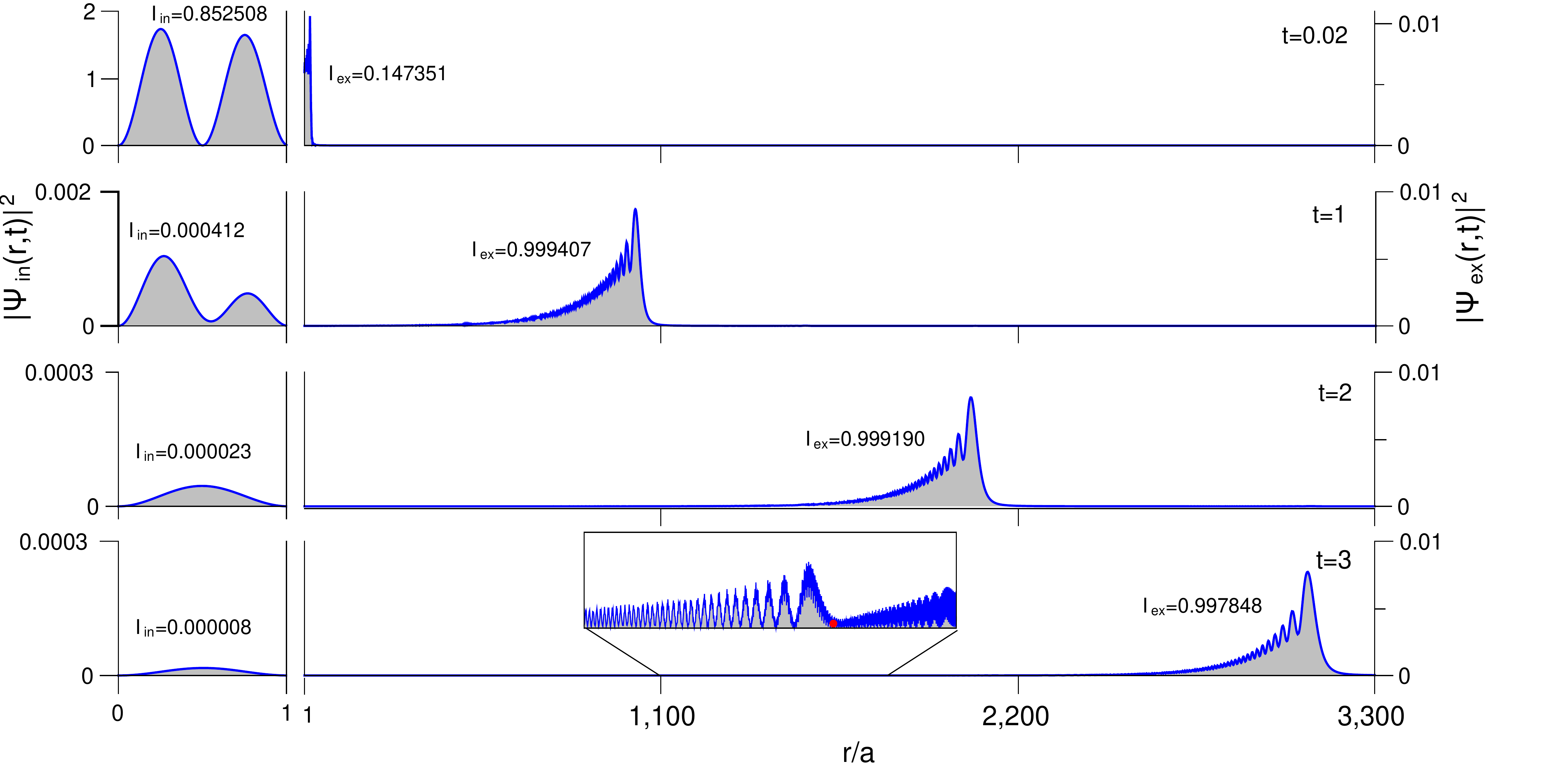}
\caption{The same as in Fig. \ref{figure1} for the normalized quantum box initial state ($q=2$). In this case the prapagating wavefront travels at a higher velocity $v_2$. A transition from the dacaying state $q=2$ to the decaying state $q=1$ can be visually appreciated in sequence of graphs on the left. A smaller structure associated to the state $q=1$ can be appreciated in the inset traveling lagged behind the main wavefront. The red dot indicates the classical position $r_1=v_1 t$.}
\label{figure4}
\end{figure*}
\section{Analysis of unitarity for the decaying wave solution}

Fulfillment of unitarity requires that the non-Hermitian evolving decaying wave function satisfies,
\begin{equation}
\int_0^{\infty}|\Psi(r,t)|^2 \,dr = 1.
\label{unt}
\end{equation}
Since the decaying wave function involves the time evolution of $\Psi_{in}(r,t)$ along the internal potential region and of $\Psi_{ex}(r,t)$ along the corresponding external region, we find convenient to write the left hand side of (\ref{unt}) as,
\begin{equation}
\int_0^{\infty}|\Psi(r,t)|^2 \,dr = I_{in}(t)+I_{ex}(t),
\label{ut}
\end{equation}
where,
\begin{equation}
I_{in}(t)=\int_0^{a}|\Psi_{in}(r,t)|^2 \,dr,
\label{ui}
\end{equation}
and
\begin{equation}
I_{ex}(t)=\int_a^{\infty}|\Psi_{ex}(r,t)|^2 \,dr,
\label{ue}
\end{equation}
with $\Psi_{in}(r,t)$ and $\Psi_{ex}(r,t)$ given, respectively, by (\ref{b6i}) and (\ref{b6e}).
At the initial time $t=0$, for an initial state normalized to unity one may write,
\begin{equation}
\int_0^{\infty}|\Psi(r,0)|^2 \,dr = I_{in}(0)=1.
\label{u0}
\end{equation}

The term $I_{in}(t)$ given by (\ref{ui}) corresponds to the nonescape probability which has been amply studied \cite{gcmv07}. For typical decaying systems one may omit the nonexponential contribution $R(r,t)$ in the expression for  $\Psi_{in}(r,t)$ (\ref{i2a}) so the decay mainly diminishes in an exponential fashion.

Let us now analyze the asymptotic behavior of the propagating decaying solution $\Psi_{ex}(r,t)$ given in (\ref{b6e}) for values $r \geq a$, that is for asymptotically large values of $r$. One sees immediately that the corresponding $r$-dependence is contained in the propagating function $M(y_n)$, whose argument $y_n$ is given by (\ref{16cd}). It follows by inspection of this expression that for a given value of the time $t$, $r \gg a$ implies
\begin{equation}
|r| \gg |2\kappa_nt|,
\label{15aa}
\end{equation}
and hence $y_n$ behaves as,
\begin{equation}
y_n \approx  \frac{1}{2}\,e^{-i\pi /4}\frac{1}{t^{1/2}}\,r.
\label{16cdz}
\end{equation}
One then may obtain, in view of the right-hand side expression in (\ref{16c}), the asymptotic expansion of $M(y_n)$ for $|y_n| \gg 1$ \cite{abramowitzchap7},
\begin{equation}
M(y_n) \approx \frac{1}{2}{\rm e}^{(ir^2/2 t)} \left [
\frac{1}{\pi^{1/2}y_n}-\frac{1}{\pi^{1/2}y_n^3}+... \right ],
\label{m6}
\end{equation}
to write the leading term of the decaying wave function $\Psi(r,t)$ for $r \gg a$ as,
\begin{equation}
\Psi_{ex}(r,t) \approx \frac{1}{\pi^{1/2}} e^{i\pi/4} e^{ir^2/4 t}t^{1/2}\,\sum_{-\infty}^{\infty} \,C_nu_n(a)\, \frac{1}{r},
\label{m6x}
\end{equation}
which shows that $\Psi_{ex}(r,t)$ is  quadratically integrable. Notice that in (\ref{m6x}), the given value of $t$ must satisfy (\ref{15aa}).

From the above considerations, one sees therefore, that as time evolves the initial probability density decays to the outside. Equation (\ref{m6x}) means from the well known expression for flux conservation \cite{cohen20} that since the probability current vanishes at infinity  one obtains,
\begin{equation}
\frac{\partial}{\partial t} \int_0^{\infty}|\Psi(r,t)|^2 \,dr =0,
\label{flux}
\end{equation}
where $\Psi(r,t)$ is given by (\ref{b6}). Since the initial state is normalized to unity, (\ref{flux}) means that at each instant of time $t$ the non-Hermitian expression (\ref{unt}) for the decaying wave satisfies unitarity, so one may write,
\begin{equation}
\int_0^{\infty}|\Psi(r,t)|^2 \,dr = I_{in}(t)+I_{ex}(t)=1.
\label{utf}
\end{equation}

Next section exemplifies our findings for an exactly solvable model.

\section{Model}
As an example, let us consider the s-wave $\delta$-shell potential, which has shown to provide an excellent  qualitative description of tunneling decay \cite{winter61,gcmv12,gcmv13},
\begin{equation}
V(r)=\lambda\delta(r-a),
\label{eq}
\end{equation}
where $\lambda$ stands for the intensity of the potential and $a$ for the radius. In our example we use $\lambda=100$ and $a=1$. The resonance solutions to the problem with complex energy eigenvalues $\kappa_n^2=\mathcal{E}_n-i\Gamma_n/2$ read,
\begin{equation}
u_n(r)=\left\{
\begin{array}{cc}
A_n \,\sin (\kappa_nr) & r\leq a \\[.4cm]
B_n \,e^{i\kappa_nr}, & r \geq a,
\end{array}
\right.
\label{5c}
\end{equation}
where we recall that $\kappa_n=\alpha_n-i\beta_n$.
From the continuity of the above solutions and the discontinuity of its derivatives with respect to $r$ (due to the $\delta$-function interaction) at the boundary value $r=a$, it is obtained that the set of $\kappa_n$'s satisfy the equation,
\begin{equation}
2i\kappa_n + \lambda ( e^{2i\kappa_na}-1)=0.
\label{5d}
\end{equation}
For $\lambda > 1$ one may write the approximate analytical solutions to Eq. (\ref{5d}) as \cite{gc10}
\begin{equation}
\kappa_n \approx \frac{n\pi}{a} \left (1-\frac{1}{\lambda a}\right ) -i \,\frac{1}{a}\left( \frac{n\pi}{\lambda a}\right )^2.
\label{5e}
\end{equation}
One may solve numerically (\ref{5d}) by using iterative procedures as the Newton-Rapshon method, which allow us to calculate the complex poles $\kappa_n$ with the desired degree of approximation using the approximate solution given by (\ref{5e}) to generate the initial values.
For a given value of the intensity $\lambda$ and a radius $a$ of the $\delta$-potential, one may then  evaluate the corresponding set of complex poles $\{\kappa_n\}$ and the set of normalized resonance states $\{u_n(r)\}$.

Since except at ultrashort times there is no memory of the initial state in the decay process \cite{cgcrv11,cgc12}, we model the initially confined state by an infinite barrier box state,
\begin{equation}
\Psi(r,0)=\sqrt{\frac{2}{a}} \, \sin \left (\frac{q\pi}{a}r \right ), \quad q=1,2,3... ,
\label{is}
\end{equation}
from which the expansion coefficients $\{C_n\}$ to the problem are easily obtained analytically.

A relevant feature of the resonance formalism is that it provides exact analytic time-dependent solutions both within the internal region of the potential and along the external region. This allow us to see the time evolution of the decay of the initial state in the internal region as well the propagation along the external region.

Figure \ref{figure1} provides a plot of $|\Psi_{G}(r,t)|^2$, given by (\ref{b6x}), and $|\Psi_{ex}(r,t)|^2$, given by (\ref{b6e}), as a function of the distance $r/a$ at the fixed time $t_0=0.5\, \tau$, with $\tau$ the lifetime of the system, to exhibit the exponential catastrophe of Gamow's solution and the propagating wavefront of the resonance expansion solution. A similar figure has been presented in Ref. \cite{dijk99}. It is worth pointing out that the resonance-state expansion for the propagating transmitted time-dependent solution of a double barrier resonant tunneling structure \cite{gcr97}, looks similar to $|\Psi_{ex}(r,t)|^2$ in Fig. \ref{figure1}.

Figure \ref{figure2} exhibits snapshots of the decay and propagation of the normalized initial state given by Eq. (\ref{is}) for $q=1$ (ground state of a quantum box). The left panel contains plots of the probability density $|\Psi_{in}(r,t)|^2$ vs $r$ in the confining internal region calculated from (\ref{b6i}), whereas the right panel exhibits the corresponding plots of the propagating probability density $|\Psi_{ex}(r,t)|^2$ vs $r$, using (\ref{b6e}), at each of the selected times $t$. The numerical values of $t$ are indicated in the upper right corner of each graph. This graphical representation allows the visualization of the time evolution of the probability density $|\Psi(r,t)|^2$ for quantum decay.  We can see that as the  probability density decreases inside the confining potential it grows and propagates along the external region  with a wavefront which is situated at approximately the classical position $r_n=v_n t$ with $v_n=2 \alpha_n$. In the present case $n=1$, which follows from the fact that the dominant term in the decay process corresponds to the coefficient $C_1^2 \approx 0.9$. We recall that the lifetime is given by $1/\Gamma_1$. The relevant point of Fig. \ref{figure2} is that at each time $t$, Eq. (\ref{ut}) is essentially satisfied.  By integrating numerically both (\ref{ui}) and (\ref{ue}), we calculate the values of $I_{in}(t)$ and $I_{ex}(t)$ indicated in each graph of fig. \ref{figure2}, and we obtain for each value of $t$, at least  $I(t)=I_{in}(t)+I_{ex}(t)= 0.999$. In order to ensure a good approximation to the exact result, in our numerical integration we choose a long enough distance (up to $r/a=4000$) along the tail on the right of the main wavefronts of Fig. \ref{figure1}. Although these tails look very smooth, they actually have a fine structure not visible in Fig. \ref{figure1}, as we show below making an appropriate zoom.

Figure \ref{figure3} shows a zoom of the second graph of Fig. \ref{figure2}. It exhibits a succession of peaks whose wavefronts correspond to the contribution to decay of resonance levels corresponding to the high resonance energy levels. The distinct red dots, represent the position of the propagating resonance terms at the positions $r_n \approx 2\alpha_n t$, with $n=2,3,4,...$. There is, in fact, an infinite number of forerunners, in agreement with the nonrelativistic character of the formalism. It is worth mentioning the complexity exhibited by these propagating structures all of which contribute to the integral term $I_{ex}(t)$ to ensure that the unitarity condition (\ref{ut}) is fulfilled.

Figure \ref{figure4}, in a similar fashion as in Fig. \ref{figure2} for the initial state $q=1$, exhibits snapshots of the decay and propagation of the normalized initial state given by Eq. (\ref{is}) for $q=2$ (second state of a quantum box). As can be seen on the left graphs of this figure, a transition from the decaying state $q=2$ (characterized by two maxima) to the decaying state $q=1$ (with a single maximum) occurs in the internal region. Initially, the decaying state is governed by the second resonance state of the system, whose main wavefront on the outside propagates at approximately the classical position $r_2= 2\alpha_2 t$ (clearly visible in each snapshot of Fig. \ref{figure4}). After the transition, the decaying state is the first resonance state of the system, and the corresponding wavefront is traveling behind the main wavefront. In view of the relatively small amplitude of the latter (since in this case $C_1^2 \ll C_2^2\approx 0.9$, we need to make a zoom on the graphs around the classical position $r_ 1= 2\alpha_1 t$ to see it. The inset on the last graph on the right of Fig.  \ref{figure4} exhibits such a traveling structure. In each snapshot shown in Fig.  \ref{figure4}, the values of the integrals $I_{in}(t)$ and $I_{ex}(t)$ are also shown, and we can verify that Eq. (\ref{ut}) is satisfactorily fulfilled.

\section{Concluding remarks}

We have shown that the exact analytical solution for the decaying wave solution, given by Eqs. (\ref{b6}), (\ref{b6i}) and (\ref{b6e}),  involving the full set of  resonant eigenfunctions and transient functions to the problem, satisfies unitarity at each instant of time.
It is worth emphasizing that the square integrability of the  decaying wave solution  follows from the fact that it behaves asymptotically at large distances as (\ref{m6x}) instead of blowing up exponentially.
Our result involving a non-Hermitian Hamiltonian obeying outgoing boundary conditions provides an exact analytical description of the tunneling decay process and favours the idea of incorporating in a fundamental fashion this non-Hermitian treatment of the Hamiltonian  to the formalism of quantum mechanics. Finally, we would like to comment that our exact result on unitarity using transient functions might be of interest in studies of quasinormal modes of black holes \cite{starinets09,hui19}.

\section*{Acknowledgements}
G.G-C. acknowledges financial support of DGAPA-UNAM-PAPIIT grants IN105216 and IN110220, Mexico; and  R.R.  acknowledges financial support from PRODEP-SEP, Mexico, under the program \textit{Apoyo para Estancias Cortas de Investigaci\'on}. R.R. also thanks Instituto de F\'{\i}sica of UNAM for its hospitality.

\section*{References}

\begin{thebibliography}{10}
\providecommand{\url}[1]{{#1}}
\providecommand{\urlprefix}{URL }
\expandafter\ifx\csname urlstyle\endcsname\relax
  \providecommand{\doi}[1]{DOI \discretionary{}{}{}#1}\else
  \providecommand{\doi}{DOI \discretionary{}{}{}\begingroup
  \urlstyle{rm}\Url}\fi

\bibitem{gamow28a}
G.~Gamow, Z. Phys. \textbf{51}, 204 (1928).
\newblock \doi{10.1007/BF01343196}

\bibitem{gamow28b}
G.~Gamow, Nature \textbf{122}, 805 (1928).
\newblock \doi{doi:10.1038/122805b0}

\bibitem{gamow49}
G.~Gamow, C.L. Critchfield, \emph{Theory of Atomic Nucleus and Nuclear
  Energy--Sources} (Oxford at the Clarendon Press, 1949)

\bibitem{sakaki87}
M.~Tsuchiya, T.~Matsusue, H.~Sakaki, Phys. Rev. Lett. \textbf{59}, 2356 (1987)

\bibitem{jochim11}
F.~Serwane, G.~Z\"urn, T.~Lompe, T.~Ottenstein, A.N. Wenz, S.~Jochim, Science
  \textbf{332}, 336 (2011).
\newblock \doi{10.1126/science.1201351}

\bibitem{bender07}
C.M. Bender, Rep. Prog. Phys. \textbf{70}, 947 (2007).
\newblock \doi{10.1088/0034-4885/70/6/R03}

\bibitem{barton63}
G.~Barton, \emph{Advanced Field Theory} (Wiley, New York, 1963).
\newblock Chap. 12

\bibitem{siegert39}
A.F.J. Siegert, Phys. Rev. \textbf{56}, 750 (1939)

\bibitem{peierls55}
R.E. Peierls, in \emph{Proceedings of the International Conference on Nuclear
  and Meson Physics}, ed. by E.J. Bellamy, R.G. Moorhouse (Pergamon Press,
  London, 1955), pp. 296--299

\bibitem{peierls59}
R.E. Peierls, Proc. Roy. Soc. A \textbf{253}, 16 (1959).
\newblock \doi{10.1098/rspa.1959.0176}

\bibitem{rosenfeld61}
J.~Humblet, L.~Rosenfeld, Nuclear Physics \textbf{26}(4), 529  (1961).
\newblock \doi{10.1016/0029-5582(61)90207-3}

\bibitem{gcp76}
G.~Garc\'{i}a-Calder\'{o}n, R.E. Peierls, Nucl. Phys. \textbf{A 265}({3}), 443
  (1976).
\newblock \doi{10.1016/0375-9474(76)90554-6}

\bibitem{madrid12}
R.~de~la Madrid, J. Math. Phys. \textbf{53}, 102113 (2012).
\newblock \doi{10.1063/1.4758925}

\bibitem{leung98}
E.S.C. Ching, P.T. Leung, A.~Maassen van~den Brink, W.M. Suen, S.S. Tong,
  K.~Young, Rev. Mod. Phys. \textbf{70}, 1545 (1998).
\newblock \doi{10.1103/RevModPhys.70.1545}

\bibitem{tolstikhin08}
O.I. Tolstikhin, Phys. Rev. A \textbf{77}, 032711 (2008).
\newblock \doi{10.1103/PhysRevA.77.032711}

\bibitem{tolstikhin98}
O.I. Tolstikhin, V.N. Ostrovsky, H.~Nakamura, Phys. Rev. A \textbf{58}, 2077
  (1998).
\newblock \doi{10.1103/PhysRevA.58.2077}

\bibitem{rosas08b}
N.~Fern\'andez-Garc\'ia, O.~Rosas-Ortiz, Ann. Phys. (N.Y:) \textbf{323}, 1397
  (2008).
\newblock \doi{10.1016/j.aop.2007.11.002}

\bibitem{moiseyev11}
N.~Moiseyev, \emph{Non-Hermitian Quantum Mechanics} (Cambridge University
  Press, 2011).
\newblock P. 5

\bibitem{khalfin58}
L.A. Khalfin, Soviet Physics JETP \textbf{6}, 1053 (1958)

\bibitem{khalfin68}
L.A. Khalfin, JETP Lett. \textbf{8}, 65 (1968)

\bibitem{chiu77}
C.B. Chiu, E.C.G. Sudarshan, B.~Misra, Phys. Rev. D \textbf{16}, 520 (1977).
\newblock \doi{10.1103/PhysRevD.16.520}

\bibitem{raizen97}
S.R. Wilkinson, C.F. Bharucha, M.C. Fischer, K.W. Madison, P.R. Morrow, Q.~Niu,
  B.~Sundaram, M.G. Raizen, Nature \textbf{387}, 575 (1997).
\newblock \doi{10.1038/42418}

\bibitem{raizen01}
M.C. Fischer, B.~Guti\'errez-Medina, M.G. Raizen, Phys. Rev. Lett. \textbf{87},
  040402 (2001).
\newblock \doi{10.1103/PhysRevLett.87.040402}.
\newblock \urlprefix\url{http://link.aps.org/doi/10.1103/PhysRevLett.87.040402}

\bibitem{monk06}
C.~Rothe, S.I. Hintschich, A.P. Monkman, Phys. Rev. Lett. \textbf{96}, 163601
  (2006)

\bibitem{torrontegui09}
E.~Torrontegui, J.G. Muga, J.~Martorell, D.W.L. Sprung, Phys.Rev. A
  \textbf{80}(1), 012703 (2009)

\bibitem{gcmv13}
G.~Garc\'{\i}a-Calder\'on, I.~Maldonado, J.~Villavicencio, Phys. Rev. A
  \textbf{88}, 052114 (2013).
\newblock \doi{10.1103/PhysRevA.88.052114}

\bibitem{mgcm05}
R.~de~la Madrid, G.~Garc\'{\i}a-Calder\'on, J.G. Muga, Czech. J. Phys.
  \textbf{55}, 1141 (2005).
\newblock Eprint arXiv: quant-ph/0512242

\bibitem{gc10}
G.~Garc\'{\i}a-Calder\'on, Adv. Quant. Chem. \textbf{60}, 407  (2010).
\newblock \doi{10.1016/S0065-3276(10)60007-X}

\bibitem{abramowitzchap7}
M.~Abramowitz, I.~Stegun, \emph{Handbook of Mathematical Functions} (Dover, N.
  Y., 1968).
\newblock Chap. 7

\bibitem{faddeyeva61}
V.N. Faddeyeva, M.N. Terentev, \emph{Tables of values of the function $
  \omega(z) = e^{-z^2} \left(1 + \frac{2i}{\sqrt{\pi}} \int_0^z e^{t^2}
  \textrm{d} t \right)$, for complex argument} (Edited by Academician V. A.
  Fock, printed in Grear Britain by Pergamon Printing \& Art Services LTD.
  London, 1961)

\bibitem{cgcm09}
A.~del Campo, G.~Garc\'{\i}a-Calder\'on, J.~Muga, Physics Reports
  \textbf{476}(1-3), 1 (2009)

\bibitem{moshinsky51}
M.~Moshinsky, Phys. Rev. \textbf{84}, 525 (1951)

\bibitem{moshinsky52}
M.~Moshinsky, Phys. Rev. \textbf{88}, 626 (1952)

\bibitem{dalibard96}
P.~Szriftgiser, M.~Guery-Odelin, M.~Arndt, J.~Dalibard, Phys. Rev. Lett.
  \textbf{77}, 4 (1996)

\bibitem{gcr97}
G.~Garc\'{i}a-Calder\'{o}n, A.~Rubio, Phys. Rev. A \textbf{55}, 3361 (1997).
\newblock \doi{10.1103/PhysRevA.55.3361}

\bibitem{rosenfeld65}
L.~Rosenfeld, Nucl. Phys. \textbf{70}, 1 (1965)

\bibitem{nussenzveig92}
H.M. Nussenzveig, \emph{Moshinsky functions, Resonances and Tunneling}
  (Springer-Verlag, 1992), chap.~19, pp. 293--310

\bibitem{dijk99}
W.~van Dijk, Y.~Nogami, Phys. Rev. Lett. \textbf{83}, 2867 (1999)

\bibitem{cgc12}
S.~Cordero, G.~Garc\'\i{}a-Calder\'on, Phys. Rev. A \textbf{86}, 062116 (2012).
\newblock \doi{10.1103/PhysRevA.86.062116}

\bibitem{gcmv07}
G.~Garc\'{\i}a-Calder\'on, I.~Maldonado, J.~Villavicencio, Phys. Rev. A
  \textbf{76}, 012103 (2007).
\newblock \doi{10.1103/PhysRevA.76.012103}

\bibitem{gcv06}
G.~Garc\'ia-Calder\'on, J.~Villavicencio, Phys. Rev. A \textbf{73}, 062115
  (2006)

\bibitem{gcr16}
G.~Garc\'{\i}a-Calder\'on, R.~Romo, Phys. Rev. A \textbf{93}, 022118 (2016).
\newblock \doi{10.1103/PhysRevA.93.022118}.
\newblock \urlprefix\url{http://link.aps.org/doi/10.1103/PhysRevA.93.022118}

\bibitem{gcch17}
G.~Garc\'{i}a-Calder\'on, L.~Chaos-Cador, Fortschritte der Physik
  \textbf{65}(6-8), 1600037 (2017).
\newblock \doi{10.1002/prop.201600037}.
\newblock 1600037

\bibitem{gcv19}
G.~Garc\'{\i}a-Calder\'on, J.~Villavicencio, Phys. Rev. A \textbf{99}, 022108
  (2019).
\newblock \doi{10.1103/PhysRevA.99.022108}

\bibitem{muga05}
A.~Ruschhaupt, F.~Delgado, J.G. Muga, J. Phys A: Math. Gen. \textbf{38}, L171
  (2005).
\newblock \doi{10.1088/0305-4470/38/9/L03}

\bibitem{mostafazadeh09c}
A.~Mostafazadeh, PRAMANA \textbf{73}, 269 (2009)

\bibitem{bender15}
L.P.H. Carl M.~Bender, Dorje C.~Brody, B.K. Meister, \emph{Non-Hermitian
  Hamiltonians in Quantum Physicss} (Springer, 2015), chap. Geometric Aspects
  of Space-Time Reflection Symmetry in Quantum Mechanics, pp. 185--199

\bibitem{barrios18}
V.M. Martinez~Alvarez, J.E. Barrios~Vargas, L.E.F. Foa~Torres, Phys. Rev. B
  \textbf{97}, 121401 (2018).
\newblock \doi{10.1103/PhysRevB.97.121401}

\bibitem{newtonchap12}
R.G. Newton, \emph{Scattering Theory of Waves and Particles}, 2nd edn. (Dover
  Publications INC., 2002).
\newblock Chap. 12

\bibitem{hatano09}
N.~Hatano, T.~Kawamoto, J.~Feinberg, Pramana \textbf{73 (3)}, 553 (2009).
\newblock
  \urlprefix\url{https://www.ias.ac.in/article/fulltext/pram/073/03/0553-0564}

\bibitem{gcmv12}
G.~Garc\'ia-Calder\'on, A.~M\'attar, J.~Villavicencio, Physica Scripta
  \textbf{T151}(T151), 014076 (2012).
\newblock \doi{10.1088/0031-8949/2012/T151/014076}

\bibitem{gcb79}
G.~Garc\'{\i}a-Calder\'on, B.~Berrondo, Lett. Nuovo Cimento \textbf{26}, 562
  (1979).
\newblock \doi{10.1007/BF02817045}

\bibitem{poppe90}
G.P.M. Poppe, C.M.J. Wijers, ACM Transactions on Mathematical Software
  \textbf{16}(1), 38 (1990)

\bibitem{gcrv07}
G.~Garc\'{\i}a-Calder\'on, R.~Romo, J.~Villavicencio, Phys. Rev. B \textbf{76},
  035340 (2007)

\bibitem{cgcrv11}
S.~Cordero, G.~Garc\'\i{}a-Calder\'on, R.~Romo, J.~Villavicencio, Phys. Rev. A
  \textbf{84}, 042118 (2011)

\bibitem{cohen20}
C.~Cohen-Tannoudji, B.~Diu, F.~Lalo\"e, \emph{Quantum Mechanics}, vol.~1, 2nd
  edn. (Wiley-VCH, 2020)

\bibitem{winter61}
R.G. Winter, Phys. Rev. \textbf{123}, 1503 (1961).
\newblock \doi{10.1103/PhysRev.123.1503}

\bibitem{starinets09}
E.~Berti, V.~Cardoso, A.O. Starinets, Class. Quantum Grav. \textbf{26}, 16301
  (2009).
\newblock \doi{10.1088/0264-9381/26/16/163001}

\bibitem{hui19}
L.~Hui, D.~Kabatb, S.S. Wong, JCAP12(2019)020 \textbf{12}, 020 (2019).
\newblock \doi{10.1088/1475-7516/2019/12/020}

\end{thebibliography}
\end{document}